\documentclass[preprint,12pt]{elsarticle}
\usepackage{newtxtext}
\usepackage{lineno}

\usepackage[bottom]{footmisc} 

\usepackage{geometry}
\usepackage{amsmath}
\usepackage{algorithm}
\usepackage{algpseudocode}
\usepackage{url}
\usepackage{ upgreek }
\usepackage{pdflscape}
\usepackage{caption}
\usepackage{setspace}
\usepackage{pifont}
\usepackage{amsmath}
\usepackage{bbding}
\usepackage{amsfonts} 
\usepackage{amssymb}
\usepackage{mathrsfs}

\usepackage{color}
\usepackage{ dsfont }
\usepackage[misc]{ifsym}

\usepackage{amsmath}      
\usepackage{graphicx}  
\usepackage{array}
\usepackage{verbatim}
\usepackage{titletoc}
\usepackage{color,xcolor}

\usepackage{units}
\usepackage{multirow}
\usepackage{tabularx}
\usepackage{booktabs}
\usepackage{bbm}
\usepackage{bm} 
\usepackage{titlesec}
\usepackage{algorithm}
\usepackage{algpseudocode}

\titleformat{\paragraph}
  {\normalfont\normalsize\itshape}{\theparagraph}{1em}{}
\titlespacing*{\paragraph}
  {0pt}{1.5ex plus 0.5ex minus .2ex}{0.5ex plus .2ex}

\usepackage{amsmath}

\usepackage{xcolor}
\definecolor{DeBlue}{RGB}{0,0,0}
\definecolor{DeGreen}{RGB}{84,130,53}
\definecolor{DeBlack}{RGB}{0,0,0}

\usepackage{natbib}
\usepackage{fancyhdr}

\usepackage{hyperref}

\begin{document}

\title{An End-to-end Building Load Forecasting Framework with Patch-based Information Fusion Network and Error-weighted Adaptive Loss}
\begin{frontmatter}

\affiliation[inst1]{organization={School of Economic and Management},
            addressline={North China Electric Power University}, 
            city={Beijing, 102206},
            country={China}}

\affiliation[inst2]{organization={School of Electrical and Electronic Engineering},
             addressline={Nanyang Technological University}, 
             city={50 Nanyang Avenue, 639798},
             country={Singapore}}

\affiliation[inst3]{organization={College of Energy and Electrical Engineering},
             addressline={Qinghai University}, 
             city={Xining, 810036},
             country={China}}

\affiliation[inst4]{organization={Department of Electrical Engineering},
             addressline={Tsinghua University},
             city={Beijing, 100084},
             country={China}}

\author[inst1]{Hang Fan}

\author[inst1]{Ying Lu} 

\author[inst2]{Weican Liu\corref{cor1}}\footnotesize
\cortext[cor1]{W. Liu is the Principal corresponding author (Email: weican001@e.ntu.edu.sg).}

\author[inst1]{Dunnan Liu} 

\author[inst3]{Xiaotao Chen} 

\author[inst3,inst4]{Shengwei Mei\corref{cor2}}
\cortext[cor2]{S. Mei is the Co-corresponding author (Email: meishengwei@tsinghua.edu.cn).}

\normalsize
\begin{abstract}
Accurate building load forecasting plays a critical role in facilitating demand response aggregation and optimizing energy management. However, the complex temporal dependencies and high volatility of building loads limit the improvement of prediction accuracy. To this end, we propose a novel end-to-end building load forecasting framework. Specifically, the framework can be divided into two main stages. In the two-stage data preprocessing module enhanced by interpretable feature selection, we utilize the Local Outlier Factor (LOF) algorithm to accurately detect and correct anomalies in the original building load series. Furthermore, we employ SVM-SHAP feature analysis to quantify the impact of environmental variables, filtering out critical feature combinations to mitigate redundancy. In the building load forecasting module, we propose the patch-based information fusion network (PIF-Net). This model applies patching technology to process input series into local blocks, extracting temporal features through a shared Gated Recurrent Unit (GRU) network with residual connections. Subsequently, an information fusion module based on a customized gating mechanism integrates the ensemble hidden states to weight the importance of different temporal patches dynamically. Additionally, the framework is trained using a novel Error-weighted Adaptive Loss (EWAL) function. By combining a rational quadratic function and logarithmic loss to dynamically adjust penalty weights based on real-time prediction error distributions, EWAL significantly enhances the model's robustness under extreme load conditions. Finally, extensive experiments demonstrate the effectiveness and superiority of our proposed framework. 
\end{abstract}


\end{frontmatter}
\setcounter{page}{1}

\section{Introduction}

\subsection{Background and motivation}

As the global energy transition accelerates and electricity market reforms deepen, building energy consumption has emerged as a major component of terminal energy usage \cite{HAN2024132582}. In this context, building load aggregators play a pivotal role in modern power systems by pooling distributed energy resources to participate effectively in the demand response (DR) market. Accurate building load forecasting not only helps load aggregators formulate optimal bidding strategies in the demand response market, but also significantly improves the operational efficiency and reliability of the entire power grid \cite{QIN2026121278}. However, building loads exhibit strong non-linearity and high volatility, especially during DR periods when energy consumption patterns shift abruptly \cite{LI2023108845}. Moreover, the fluctuation of building loads is strongly influenced by multiple interactive environmental factors \cite{RAZA20151352}. Effectively identifying the most critical environmental drivers while mitigating the interference of noisy data and feature redundancy remains a significant challenge. To this end, developing an end-to-end forecasting framework with strong adaptive capabilities and effective feature fusion holds substantial significance.

Currently, mainstream building load forecasting models tend to input long sequences directly as a whole. However, this processing paradigm is prone to information loss when capturing long-range dependencies and often overlooks local, fine-grained evolutionary patterns. Regarding feature selection, commonly used methods include filtering methods, wrapper methods, and embedding methods \cite{NEUBAUER2024122668}. Nevertheless, these approaches struggle to effectively identify hidden local anomalies within time series and lack the capacity for interpretable quantification of the impacts exerted by complex environmental variables \cite{WANG2025111910}. In terms of loss function design, existing studies predominantly employ static metrics, such as Mean Squared Error (MSE) or Mean Absolute Error (MAE) as optimization objectives \cite{ZHANG2017270,ZHENG2026120983}. Under extreme fluctuation scenarios like demand response events, these static loss functions are overly sensitive to outliers or abrupt change points, rendering the models susceptible to falling into local optima. To address the aforementioned limitations, this paper proposes an end-to-end building load forecasting framework (PIF-Net). By leveraging patch-based information, this framework effectively extracts and dynamically fuses local and global spatiotemporal dependencies. Furthermore, an Error-Weighted Adaptive Loss is introduced to dynamically adjust penalty weights based on the real-time prediction error distribution during training, thereby significantly enhancing the model's robustness and forecasting stability under extreme load conditions.

\subsection{Related work}
Research on building load forecasting generally falls into four categories: physical modeling, statistical, shallow machine learning, and deep learning methods. Based on thermodynamics and heat transfer principles, physical modeling methods offer clear advantages for single, specific buildings. For example, Qiang \textit{et al.} proposed a demand-side prediction framework that uses dynamic and steady-state white-box models to forecast heating loads in terminal buildings by accounting for indoor temperature variations. Their comparative analysis showed that the steady-state model provides computational benefits while estimating heating demands and validating potential energy savings of 11\%--27\% through indoor temperature optimization \cite{zhang2020development}. Pachano \textit{et al.} developed an optimization-based calibration methodology for white-box energy models, using a genetic algorithm to fine-tune HVAC parameters. Using high-resolution operational data, this approach reduced the energy performance gap and met ASHRAE and EVO standards during both training and extended testing periods \cite{PACHANO2021111380}. Furthermore, Yu \textit{et al.} introduced a community-level load forecasting method that accounts for inter-building interactions. Their approach establishes a Community Occupant Agent Model (COAM) for occupant data generation and uses AnyLogic to simulate decision-making behaviors. This framework captures occupant-building dynamics to assess community heating loads \cite{yu2023bottom}.

Despite these advantages, physical modeling methods face limitations in practice, including complex model calibration, high costs for acquiring input parameters, and difficulty in modeling human behaviors \cite{CHEN20222656}. As a result, data-driven time-series modeling approaches have gained prominence in building load forecasting \cite{PACHANO2025115485}. Statistical methods, such as AR, ARIMA, and various regression models, are widely used in short-term load forecasting \cite{DEB2017902}. For example, Norizan \textit{et al.} designed a double seasonal autoregressive integrated moving average (SARIMA) model to predict half-hourly short-term load demand in Malaysia. Compared to a standalone ARIMA model, their method decreased the in-sample and out-of-sample Mean Absolute Percentage Error (MAPE) by 10.22676\% and 57.03126\%, respectively \cite{Mohamed2011223}. Furthermore, Yang \textit{et al.} integrated an ARIMA model with a Projection Pursuit Regression (PPR) model for half-hourly load predictions, showing that the combined model outperforms standalone variants across multiple evaluation metrics \cite{8881315}. In addition, Kychkin \textit{et al.} introduced a Seasonal Persistence Regression (SPR) model, and experiments on the electrical loads of four residential buildings showed that it improves forecasting accuracy over classical statistical models \cite{KYCHKIN2021111200}.

However, traditional statistical methods often struggle to capture the nonlinearity, stochasticity, and complex temporal dependencies in building load profiles \cite{pallonetto2022forecast}. In contrast, shallow machine learning techniques perform well when processing high-dimensional features and non-stationary data \cite{kuster2017electrical}. For example, Chen \textit{et al.} introduced an SVR-based forecasting model to establish short-term load baselines for demand response in office buildings, improving prediction accuracy and stability compared to seven traditional baseline models \cite{CHEN2017659}. Chaganti \textit{et al.} developed a data-driven ensemble framework that uses three random forest algorithms (3RF). By analyzing architectural parameters, such as relative compactness, surface area, and wall area, their model forecasted heating and cooling loads in smart buildings. This approach outperformed existing machine learning models, achieving prediction accuracies of 0.999 for heating and 0.997 for cooling \cite{chaganti2022building}. In addition, Le \textit{et al.} designed a hybrid PSO-XGBoost model, combining particle swarm optimization with an extreme gradient boosting machine to estimate building heating loads for smart city planning. Their experiments showed that this hybrid model outperformed classical machine learning baselines across various statistical metrics and identified overall height, roof area, wall area, and surface area as key predictive features \cite{le2019estimating}.

Deep learning has also been widely adopted in building load forecasting because neural networks can effectively model complex dependencies within load profiles \cite{FAN2026131620, FAN2026125296}. For example, Waseem \textit{et al.} compared the performance of an artificial neural network (ANN) against a random forest (RF) model for building energy consumption forecasting. Results from a hotel in Madrid, Spain, showed that the ANN achieved higher predictive accuracy than the RF model \cite{ahmad2017trees}. Guo \textit{et al.} investigated the performance of three prevalent forecasting methods: SVM, RF, and LSTM. Their results indicated that LSTM achieved the highest accuracy, reducing the Root Mean Square Error (RMSE) by 86.1863 and 19.10172 compared to SVM and RF, respectively \cite{guo2021machine}. In addition, L’Heureux \textit{et al.} designed a transformer-based architecture for electrical load forecasting that adapts the natural language processing workflow by adding N-space transformation and a technique for processing contextual features. Evaluations across diverse data streams and forecasting horizons showed that this approach effectively manages contextual time-series data, outperforming sequence-to-sequence (Seq2Seq) models \cite{l2022transformer}. Furthermore, Neubauer \textit{et al.} introduced a transfer learning framework for short-term heating load forecasting by combining SHAP-based feature importance with load pattern metrics. Validated across datasets encompassing six distinct building archetypes, including residential, hotel, and public buildings, their study highlighted the performance of the PatchTST model due to its lightweight architecture and high precision \cite{neubauer2026transfer}. Qin and Evins developed a deep temporal convolutional residual neural network (ResNet) to predict 24-hour heating load profiles at a 15-minute resolution. Using a representative weather sampling scheme, this approach achieved high predictive accuracy with an $R^2$ of up to 0.9978. It effectively preserved key demand response indicators while enabling computationally efficient evaluations at both building and district scales \cite{qin2026deep}. \textbf{Table \ref{tab1}} contrasts our work with the literature.

\begin{table}[h]
\centering
\small
\caption{\textbf{Contrasting our work with the literature}}
\label{tab1}
\setlength{\tabcolsep}{2.7mm}
\begin{tabular}{ccccccc} 
\toprule
\multicolumn{2}{c}{Reference} & \begin{tabular}[c]{@{}c@{}}Data \\ Processing\end{tabular} & \begin{tabular}[c]{@{}c@{}}Feature \\ Selection\end{tabular} & \begin{tabular}[c]{@{}c@{}}Patch \\ Processing\end{tabular} & \begin{tabular}[c]{@{}c@{}}Gating \\ Mechanism\end{tabular} & \begin{tabular}[c]{@{}c@{}}Loss\\ function\end{tabular} \\ 
\midrule
\multirow{2}{*}{\textit{Physical Model}} & \citep{zhang2020development} & \checkmark & \checkmark & \ding{55} & \ding{55} & \checkmark \\
 & \citep{PACHANO2021111380} & \checkmark & \checkmark & \ding{55} & \ding{55} & \checkmark \\ \addlinespace
\multirow{2}{*}{\textit{Statistical Model}} & \citep{Mohamed2011223} & \checkmark & \ding{55} & \ding{55} & \ding{55} & \checkmark \\
 & \citep{KYCHKIN2021111200} & \ding{55} & \checkmark & \ding{55} & \ding{55} & \checkmark \\ \addlinespace
\multirow{2}{*}{\textit{Shallow ML Model}} & \citep{CHEN2017659} & \ding{55} & \checkmark & \ding{55} & \ding{55} & \checkmark \\
 & \citep{chaganti2022building} & \ding{55} & \ding{55} & \ding{55} & \ding{55} & \ding{55} \\ \addlinespace
\multirow{3}{*}{\textit{Deep Learning Model}} & \citep{neubauer2026transfer} & \ding{55} & \checkmark & \checkmark & \checkmark & \checkmark \\
 & \citep{qin2026deep} & \ding{55} & \checkmark & \ding{55} & \ding{55} & \checkmark \\
 & \textbf{Our Works} & \textbf{\checkmark} & \textbf{\checkmark} & \textbf{\checkmark} & \textbf{\checkmark} & \textbf{\checkmark} \\ 
\bottomrule
\end{tabular}
\end{table}

\subsection{Our contributions}

In this paper, we propose an end-to-end building load forecasting framework with a patch-based information fusion network and error-weighted adaptive loss. To further enhance predictive accuracy, we introduce a two-stage data preprocessing module reinforced by interpretable feature fusion. This framework provides robust technical support for building aggregator scheduling in electricity markets. Thus, the major contributions of this paper can be summarized as follows: 

\begin{itemize}
\item 
\emph{\textbf{An end-to-end building load forecasting framework with patch-based information fusion network}}: To effectively capture the complex local and global temporal dependencies in building loads, we propose PIF-Net. Specifically, we apply patching technology to process the input series into local blocks, extracting temporal features through a shared GRU network with residual connections. Then, we introduce an information fusion module on a customized gating mechanism to integrate the hidden states of the ensemble. This allows the model to dynamically weight the importance of different temporal patches, reflecting the underlying patterns of building energy consumption. 

\item 
\emph{\textbf{A two-stage data preprocessing module enhanced by interpretable feature selection}}: To reduce the negative impact of data outliers and identify the most influential environmental factors, we designed A two-stage data preprocessing module enhanced by interpretable feature selection. First, we utilize the Local Outlier Factor (LOF) algorithm to accurately detect anomalies in the original building load series. Second, we employ SVM-SHAP feature analysis to quantify the impact of various variables on load fluctuations, filtering out the most critical feature combinations to mitigate redundancy and strengthen model generalization.

\item 
\emph{\textbf{An error-weighted adaptive loss function for robust load forecasting}}: To handle the high volatility and uncertainty of building loads during demand response periods, we propose the EWAL function. Unlike traditional MSELoss, EWAL dynamically adjusts penalty weights based on the real-time prediction error distribution during training. This mechanism significantly enhances the model's stability under extreme load conditions and improves the overall precision of the load forecasting.

\end{itemize}

The rest of the paper proceeds as follows: Section 2 introduces the overall design of the PIF-Net framework. Section 3 provides an overview of our general methodology, detailing the two-stage data preprocessing module enhanced by interpretable feature fusion, the building load forecasting module, and the error-weighted adaptive loss function. Section 4 discusses data selection, parameter design, evaluation metrics, and the operating environment. Section 5 presents a performance comparison of the proposed end-to-end building load forecasting framework with other models through four experiments, while also exploring the necessity of each component of the model and its sensitivity. Finally, Section 6 concludes the study and provides prospects for future work.

\section{Problem Definition and Overall Framework}

\subsection{Problem definition}
Building load forecasting essentially belongs to the category of multivariate time series forecasting. Given a historical observation sequence composed of building loads and multiple related environmental variables, the objective is to predict the building load values for a specific future period. Mathematically, let $X = [\mathbf{x}_1, \mathbf{x}_2, \cdots, \mathbf{x}_L] \in \mathbb{R}^{L \times D}$ represent the multivariate input sequence, where $L$ denotes the length of the look-back window and $D$ is the number of input features. Let $Y = [y_{L+1}, y_{L+2}, \cdots, y_{L+T}] \in \mathbb{R}^{T}$ denote the actual building load sequence to be predicted, where $T$ is the prediction horizon. The core task of our proposed model is to learn a non-linear mapping function $\mathcal{F}(\cdot)$ parameterized by $\Theta$ to generate the predicted sequence $\hat{Y} \in \mathbb{R}^{T}$, which can be formulated as:

\begin{equation}
\begin{aligned}
&\hat{Y} = \mathcal{F}(X; \Theta),  \\
\end{aligned}
\end{equation}

The optimization goal is to find the optimal parameters $\Theta^*$ that minimize the discrepancy between the predicted sequence $\hat{Y}$ and the ground truth $Y$ over the training dataset, ensuring robust performance even under highly volatile demand response conditions.

\subsection{The whole framework design}

\begin{figure}[!t]
    \centering
    \captionsetup{labelfont=bf}
    \includegraphics[width=1.0\linewidth]{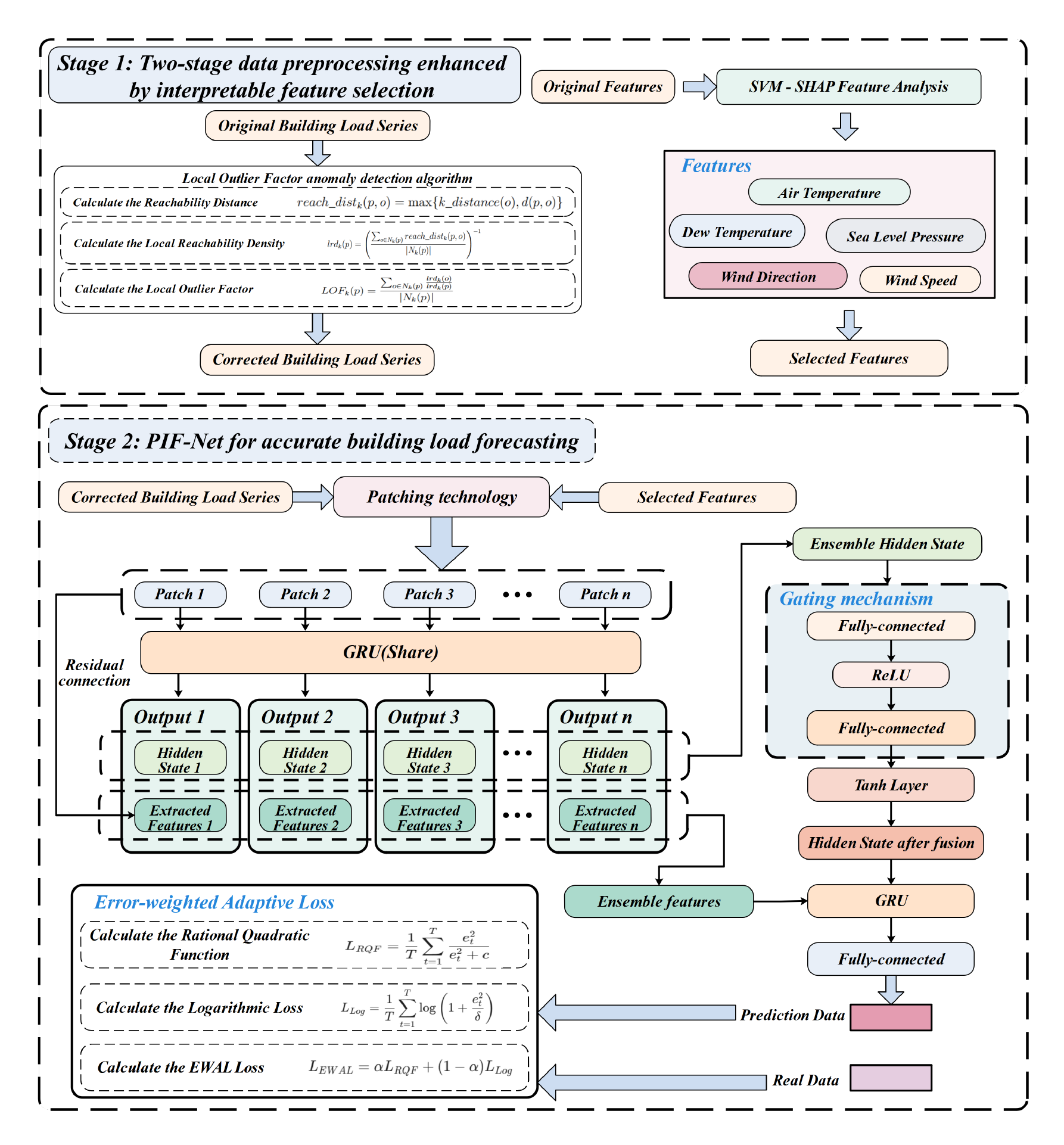}
    \vspace{-1.4\baselineskip}
    \caption{\textbf{Flow chart of PIF-Net framework}}
    \label{fig1}
\end{figure}

In this section, we introduce the proposed whole framework design. Our proposed PIF-Net framework is divided into primarily three modules: a two-stage data preprocessing module enhanced by interpretable feature selection, an accurate building load forecasting module and a customized loss function module. For the two-stage data preprocessing module enhanced by interpretable feature selection, we apply the Local Outlier Factor (LOF) algorithm to identify and correct anomalies in the raw building load series using distance metrics and local densities. We then use SVM-SHAP feature analysis to select the most relevant meteorological variables from air temperature, dew temperature, sea level pressure, wind direction, and wind speed. For the building load forecasting module, we segment the corrected load series and selected features into distinct patches. These patches are processed by a shared Gated Recurrent Unit (GRU) with a residual connection to capture temporal dynamics, yielding feature representations and hidden states for each patch. The hidden states are aggregated using a gating mechanism composed of fully connected layers, a ReLU activation, and a Tanh layer. We merge this fused hidden state with the patch features and pass them through a second GRU, generating the final load prediction via a fully connected layer. For the loss function module, we propose an Error-weighted Adaptive Loss (EWAL) function. By introducing an adaptive weight $\alpha$, EWAL dynamically balances a Rational Quadratic Function and a Logarithmic Loss. This weighting handles varying error magnitudes and improves overall forecasting robustness. \textbf{Figure \ref{fig1}} demonstrates the flow chart of our PIF-Net framework. 

\section{Methodology}

In this section, we first introduce the two-stage data preprocessing module enhanced by interpretable feature selection, which includes load data anomaly detection based on the Local Outlier Factor (LOF) algorithm, as well as feature selection with the SVM-SHAP feature analysis. Then, we provide a detailed description of the PIF-Net model used for building load forecasting and the Error-weighted Adaptive Loss (EWAL) function.

\subsection{Two-stage data preprocessing module enhanced by interpretable feature selection}

\begin{figure}[!h]
    \centering
    \captionsetup{labelfont=bf}
    \includegraphics[width=1.0\linewidth]{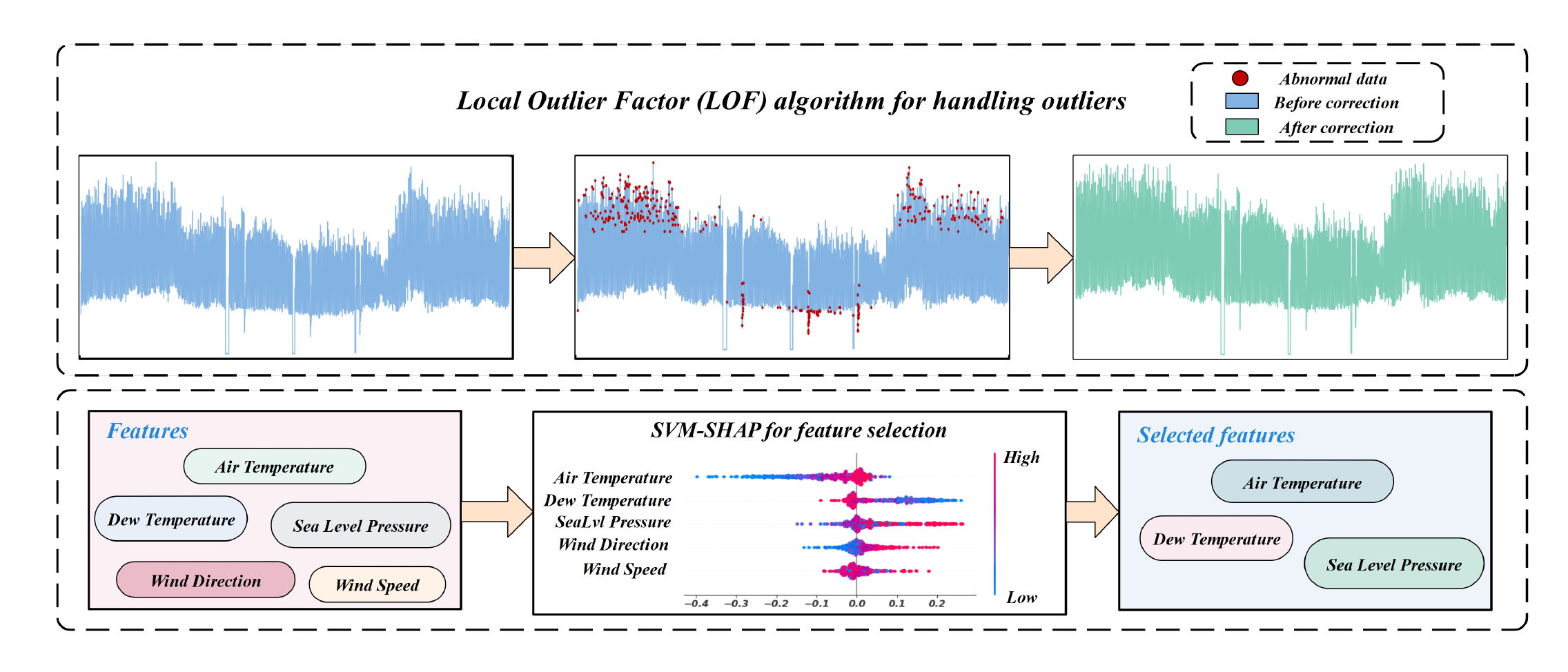}
    \vspace{-0.2\baselineskip}
    \caption{\textbf{Flow chart of the two-stage data preprocessing module enhanced by interpretable feature selection}}
    \label{fig2}
\end{figure}

Due to extreme weather, human activities, or sensor failures, building load series can experience abnormal fluctuations. To address this issue, effective data correction can help the forecasting model learn more accurate load change patterns, thereby improving the model's generalization ability. Moreover, building load data are influenced by various complex meteorological factors, so filtering out representative features can enhance the model's computational efficiency and predictive accuracy. In this section, we utilize the Local Outlier Factor (LOF) algorithm to correct outliers in building load data and employ SVM-SHAP feature analysis to select relevant meteorological features. In the following, we will provide a detailed description of the two-stage data preprocessing module enhanced by interpretable feature selection. The specific workflow is shown in \textbf{Figure \ref{fig2}}.

(1) \textbf{The Local Outlier Factor (LOF) algorithm for anomaly detection:} To ensure the quality and reliability of building load data and to reduce the negative impact of outliers on model forecasting performance, we first employ the Local Outlier Factor (LOF) algorithm to accurately detect outliers in the original building load series. Subsequently, we use the moving average method to correct these outliers, resulting in the corrected building load series.

(2) \textbf{SVM-SHAP for interpretable feature selection:} To identify key environmental features influencing building loads, reduce redundancy, and improve model generalization, we trained a Support Vector Machine (SVM) to capture the non-linear relationships between environmental variables and building loads. We then applied SHAP to quantify the importance of the features and select the optimal subset of features.

\subsubsection{Local Outlier Factor (LOF) algorithm}

To ensure the quality and reliability of the building load data, we use the Local Outlier Factor (LOF) algorithm to identify and remove outliers in the data. The LOF algorithm is a density-based anomaly detection algorithm proposed by Breunig et al. in 2000 \cite{breunig2000lof}. Specifically, it evaluates the local density of a data point by comparing it with the local densities of its $k$-nearest neighbors to continuously quantify the deviation of the dataset until each data point is assigned an anomaly score. The LOF method is based on the assumption that normal data points generally reside in neighborhoods with similar or higher densities, whereas anomaly data points are relatively isolated in the feature space and exhibit significantly lower local densities than their neighbors.

Therefore, by calculating the local density deviation of each data point, we can assess its degree of abnormality. The lower the local reachability density compared to its neighbors, the more likely it is to be an anomaly. For each data point, the calculation formula for the reachability distance is as follows:

\begin{equation}
\begin{aligned}
&reach\_dist_k(p, o) = \max\{k\_distance(o), d(p, o)\},  \\
\end{aligned}
\end{equation}
where $k\_distance(o)$ represents the distance from point $o$ to its $k$-th nearest neighbor, $d(p, o)$ represents the actual Euclidean distance between data point $p$ and point $o$. Next, the formula for the local reachability density is as follows:

\begin{equation}
\begin{aligned}
&lrd_k(p) = \left( \frac{\sum_{o \in N_k(p)} reach\_dist_k(p, o)}{|N_k(p)|} \right)^{-1} \\
\end{aligned}
\end{equation}
where $N_k(p)$ denotes the set of $k$-nearest neighbors of the data point $p$ in question. Therefore, the formula for calculating the anomaly score is as follows:

\begin{equation}
\begin{aligned}
&LOF_k(p) = \frac{\sum_{o \in N_k(p)} \frac{lrd_k(o)}{lrd_k(p)}}{|N_k(p)|} \\
\end{aligned}
\end{equation}

In addition, we correct the identified outliers using a moving average. By averaging the adjacent data points, the corrected value $p_i$ is obtained as:

\begin{equation}
\begin{aligned}
&p_i = \frac{(p_{i-1} + p_{i+1})}{2} \\
\end{aligned}
\end{equation}

\subsubsection{SVM-SHAP feature analysis}

In this part, we use the SVM-SHAP feature analysis mechanism to eliminate redundant environmental variables to improve the generalization ability of the model. SHAP is a unified framework for interpreting machine learning models, proposed by Lundberg and Lee in 2017 based on cooperative game theory \cite{lundberg2017unified}. The core idea of SVM-SHAP is to evaluate the importance of features by quantifying their marginal contributions to the model's output. Specifically, we first train a Support Vector Machine (SVM) regression model to capture the complex non-linear mappings between multidimensional environmental factors and building loads. Then, the algorithm calculates the Shapley value for each feature by considering all possible combinations of feature subsets. If a feature significantly and consistently alters the prediction output when added to various feature coalitions, its weight will be high. The formula for calculating the SHAP value of the $i$-th feature is as follows:

\begin{equation}
\phi_i = \sum_{S \subseteq F \setminus \{i\}} \frac{|S|! (|F| - |S| - 1)!}{|F|!} [\mathcal{M}_{S \cup \{i\}}(\mathbf{x}_{S \cup \{i\}}) - \mathcal{M}_S(\mathbf{x}_S)],
\end{equation}
where $\phi_i$ signifies the SHAP value (contribution) of feature $i$, $F$ represents the complete set of all features, and $S$ is a subset of features that does not contain $i$. $|F|$ and $|S|$ denote the number of features in sets $F$ and $S$, respectively. $\mathcal{M}_{S \cup \{i\}}(\mathbf{x}_{S \cup \{i\}})$ represents the model's prediction output when feature $i$ is included in the feature subset $S$, and $\mathcal{M}_S(\mathbf{x}_S)$ denotes the prediction output without feature $i$.

Furthermore, to evaluate the global impact of each environmental factor on the building load across the entire dataset, we calculate the global feature importance. Based on this metric, variables with minor impacts are eliminated to mitigate redundancy. The calculation formula for the global importance of feature $i$ is as follows:

\begin{equation}
I_i = \frac{1}{N} \sum_{j=1}^{N} |\phi_i^{(j)}|,
\end{equation}
where $I_i$ represents the global importance of feature $i$, $N$ is the total number of samples in the dataset, and $\phi_i^{(j)}$ is the SHAP value of feature $i$ for the $j$-th sample. 

\subsection{PIF-Net model for accurate building load forecasting}

\begin{figure}[!h]
    \centering
    \captionsetup{labelfont=bf}
    \includegraphics[width=1.0\linewidth]{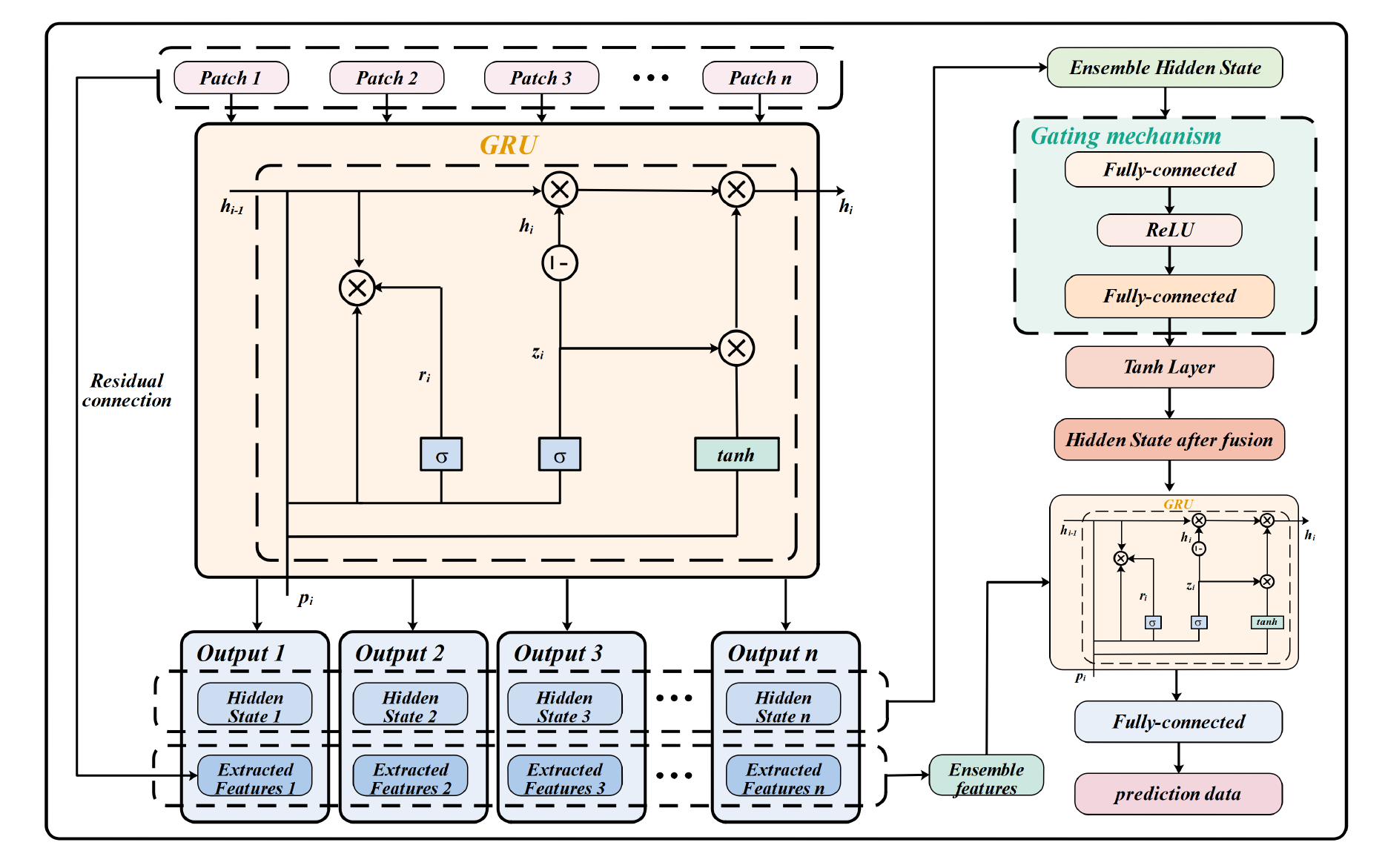}
    \vspace{-0.1\baselineskip}
    \caption{\textbf{Structure of PIF-Net model}}
    \label{fig3}
\end{figure}

In this section, we provide a detailed introduction to the proposed PIF-Net building load forecasting model. First, we utilize patching technology to partition the input multidimensional time series into multiple local blocks. Then, we adopt a shared Gated Recurrent Unit (GRU) network with residual connections to extract temporal dynamic features from these local patches. Subsequently, we use an information fusion module based on a customized gating mechanism to dynamically evaluate and fuse the extracted hidden states into a comprehensive global representation. Finally, we apply a linear projection layer to the fused data to directly map the high-dimensional features into the target prediction horizon, further ensuring the model's forecasting accuracy. \textbf{Figure \ref{fig3}} illustrates the structure of the PIF-Net model.

\subsubsection{Patching technology for local information extraction}

To efficiently extract local fine-grained temporal patterns and overcome the computational bottlenecks associated with processing long sequence data, we utilize patching technology in the forecasting model. The core idea of patching is to segment a continuous long time series into multiple sub-series patches rather than processing the data point-wise. Therefore, the model can preserve the local semantic context of the time series, allowing the subsequent network to better comprehend localized energy consumption behaviors. Specifically, given the preprocessed multivariate input sequence $X \in \mathbb{R}^{L \times D}$, the patching operation slides a window of length $P_L$ across the sequence with a stride of $S$. The total number of generated patches is calculated with the following formula:

\begin{equation}
\begin{aligned}
&P = \left\lfloor \frac{L - P_L}{S} \right\rfloor + 1, \\
\end{aligned}
\end{equation}
where $P$ indicates the total number of segmented patches, $P_L$ signifies the length of each local patch, $S$ refers to the sliding stride between adjacent patches. Then, the $i$-th extracted local patch can be calculated with the following formula:

\begin{equation}
\begin{aligned}
&\mathbf{p}_i = X[(i-1)S : (i-1)S + P_L, :], \quad i \in \{1, 2, \cdots, P\}, \\
\end{aligned}
\end{equation}
where $\mathbf{p}_i \in \mathbb{R}^{P_L \times D}$ represents the $i$-th extracted local patch block, $\lfloor \cdot \rfloor$ denotes the floor function.

Furthermore, through this patching mechanism, the original 2D input tensor $X \in \mathbb{R}^{L \times D}$ is seamlessly transformed into a 3D patch tensor $X_{patch} \in \mathbb{R}^{P \times P_L \times D}$. This spatial transformation significantly shortens the effective sequence length fed into the subsequent recurrent neural network. Consequently, it fundamentally alleviates the issues of information loss and gradient vanishing commonly encountered in traditional deep learning models when handling long-range dependencies, thereby strengthening the efficiency and stability of the subsequent feature extraction.

\subsubsection{Shared GRU network with residual connections}

To extract local temporal dynamic features from the patches while mitigating the vanishing gradient problem, we introduce a shared Gated Recurrent Unit (GRU) network with residual connections. Applying a single weight-shared GRU network independently across all segmented temporal patches reduces the number of trainable parameters and ensures the model learns consistent temporal patterns across different time windows. A residual connection adds the raw input patch to the GRU's non-linear output, preserving the original high-frequency fluctuations of the building load.

For the $i$-th segmented patch $\mathbf{p}_i \in \mathbb{R}^{P_L \times D}$, the shared GRU processes its internal elements sequentially to generate a localized hidden representation. The hidden state vector $\mathbf{h}_i$ is computed as: 

\begin{equation}
\begin{aligned}
&\mathbf{h}_i = \text{GRU}(\mathbf{p}_i; \theta_{GRU}),\ \\
\end{aligned}
\end{equation}
where $\theta_{GRU}$ denotes the shared trainable parameters of the GRU network. A residual connection then fuses the raw input with the extracted features. Because the dimensions of the flattened raw patch and the GRU output may differ, a linear projection is used for structural alignment. The enhanced representation $\mathbf{u}_i$ for the $i$-th patch is:

\begin{equation}
\begin{aligned}
&\mathbf{u}_i = \mathbf{h}_i + \mathbf{W}_{res} \text{Flatten}(\mathbf{p}_i) + \mathbf{b}_{res},\ \\
\end{aligned}
\end{equation}
where $\text{Flatten}(\mathbf{p}_i)$ flattens the 2D patch matrix $\mathbf{p}_i$ into a 1D vector. $\mathbf{W}{res}$ and $\mathbf{b}_{res}$ are the weight matrix and bias vector of the linear projection layer, respectively, used for dimension matching.

\subsubsection{Gating mechanism for information fusion}

In order to dynamically evaluate the importance of different temporal patches and aggregate them into a global representation, we design a gating mechanism for information fusion. Historical time patches contribute differently to future building load fluctuations; for example, energy consumption patterns during morning peak hours typically hold more predictive value than those during late-night off-peak hours. The gating mechanism assigns a normalized weight to the hidden state of each patch based on its informational value and computes a weighted sum to capture the global spatiotemporal context.

Given the enhanced representation sequence of all patches $U = [\mathbf{u}_1, \mathbf{u}_2, \cdots, \mathbf{u}_P]$ from the residual GRU network, the gating score $e_i$ for the $i$-th patch is computed as:

\begin{equation}
\begin{aligned}
&e_i = \mathbf{W}_{g} \mathbf{u}_i + \mathbf{b}_{g},\\
\end{aligned}
\end{equation}
where $\mathbf{W}{g}$ and $\mathbf{b}_{g}$ are the learnable weight matrix and bias vector of the gating layer. Then, a softmax function then converts these raw scores into normalized attention weights:

\begin{equation}
\begin{aligned}
&\alpha_i = \frac{\exp(e_i)}{\sum_{j=1}^{P} \exp(e_j)},\\
\end{aligned}
\end{equation}
where $\alpha_i \in (0, 1)$ signifies the normalized attention weight, satisfying $\sum_{i=1}^{P} \alpha_i = 1$. Finally, the global fused representation is obtained by computing the weighted sum of all patch representations. The formula is as follows:

\begin{equation}
\begin{aligned}
&\mathbf{C} = \sum_{i=1}^{P} \alpha_i \mathbf{u}_i,\\
\end{aligned}
\end{equation}
where $\mathbf{C}$ is the final high-dimensional global feature representation.

\subsection{Error-weighted adaptive loss function}

Traditional loss functions, such as MSE or MAE, are sensitive to sudden load spikes and prone to gradient explosion. To address these limitations and improve robustness under extreme load conditions during demand response periods, we propose an Error-weighted adaptive loss (EWAL) function. EWAL calculates an adaptive weight based on the error distribution of the current training batch. For normal load samples with small errors, the loss emphasizes the Rational Quadratic (RQ) component to maintain high fitting precision. For anomalous spikes with large errors, the weight shifts towards the Log component, smoothing the penalty to prevent unstable updates. The prediction error $e_t$ at time step $t$ is:

\begin{equation}
e_t = y_t - \hat{y}_t , 
\end{equation}
where $e_t$ represents the real-time prediction error, $y_t$ and $\hat{y}_t$ are the true and predicted building loads at time step $t$, respectively. The Rational Quadratic loss $\mathcal{L}_{RQ}$ and the Logarithmic loss $\mathcal{L}_{Log}$ are defined as:

\begin{equation}
\mathcal{L}_{RQ}(e_t) = \frac{e_t^2}{1 + (e_t / c)^2},
\end{equation}
\begin{equation}
\mathcal{L}_{Log}(e_t) = \log(1 + \frac{|e_t|}{c}),
\end{equation}
where $c$ is a hyperparameter that controls the scale of the error penalty. To balance these two components, an adaptive weight $\omega_t$ is calculated based on the standard deviation of the prediction errors in the current batch:

\begin{equation}
\omega_t = \exp\left(-\frac{|e_t|}{\sigma + \epsilon}\right),
\end{equation}
where $\sigma$ denotes the standard deviation of the errors within the current training batch, reflecting the real-time error distribution. $\epsilon$ is a small constant added to prevent division by zero. The proposed EWAL function is formulated as follows:

\begin{equation}
\mathcal{L}_{EWAL} = \frac{1}{T} \sum_{t=1}^{T} \left[ \omega_t \mathcal{L}_{RQ}(e_t) + (1 - \omega_t) \mathcal{L}_{Log}(e_t) \right].
\end{equation}

\section{Data Selection, Parameter Design, and Evaluation Index}

In this section, we introduce two building load datasets from different functional buildings, the hyperparameters and internal parameters of the two-stage data preprocessing module, the proposed PIF-Net forecasting model, and the evaluation metrics utilized for a comprehensive assessment.

\subsection{Data selection}
\begin{figure}[!h]
    \centering
    \captionsetup{labelfont=bf}
    \includegraphics[width=0.85\linewidth]{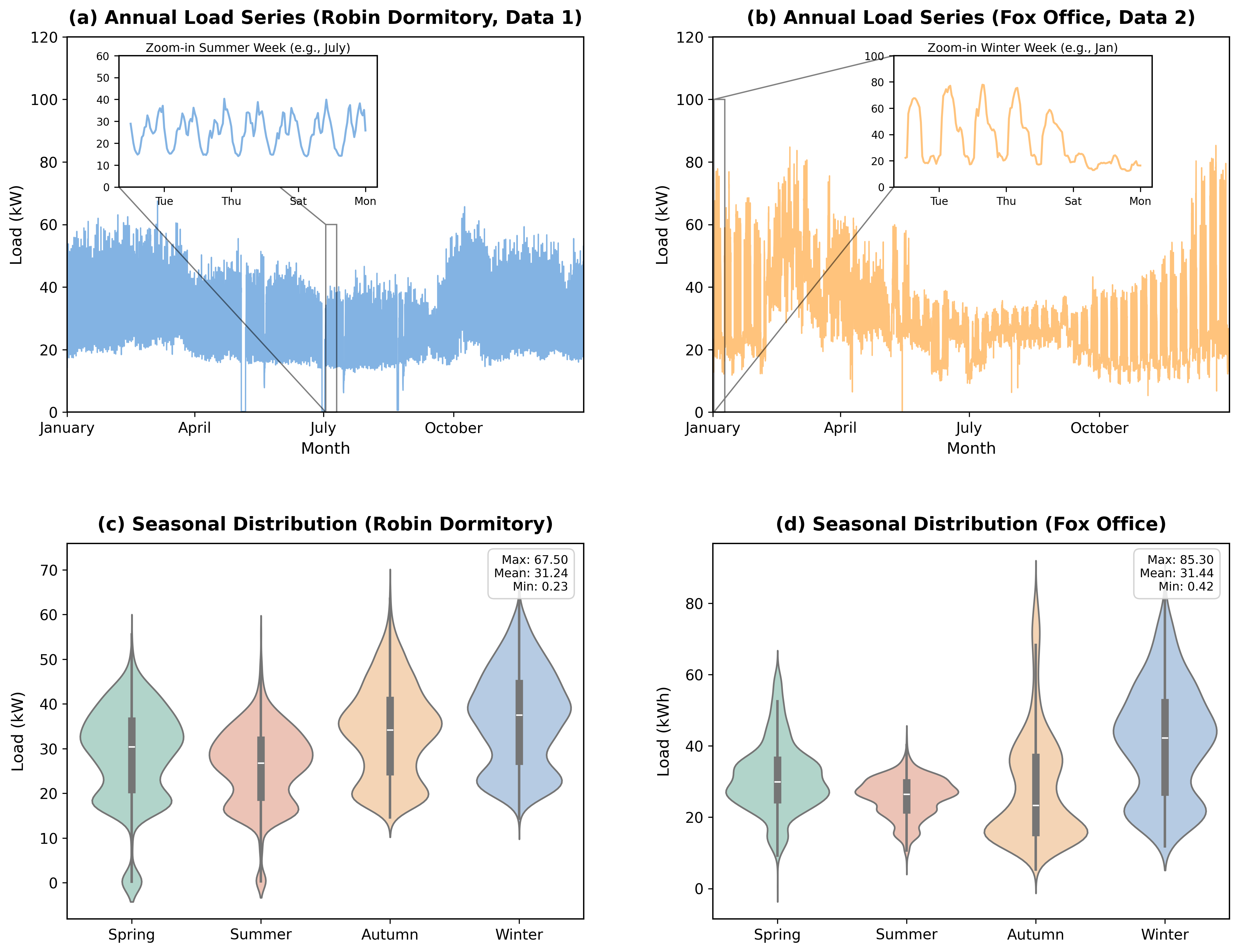}
    \vspace{-0.2\baselineskip}
    \caption{\textbf{The characteristics of the two different datasets}}
    \label{fig4}
\end{figure}

In this study, we use the Building Data Genome 2 (BDG2) dataset, which contains electricity, heating, and cooling metering data for 1,636 buildings \cite{miller2020building}. To evaluate the generalization and robustness of the proposed framework across different building functions and energy behaviors, we selected electricity load data from two distinct building types in 2017. Each subset comprises 8,760 hourly data points (8,660 for training and 100 for testing) measured in kW. \textbf{Figure \ref{fig4}} and \textbf{Table \ref{tab2}} detail the basic characteristics and statistical distributions of these datasets, highlighting their complex volatility and non-linearity.

\begin{table}[!h]
\small
\begin{spacing}{1.2}
\caption{\textbf{Basic characteristics of the different building datasets}}
\label{tab2}
\small
\setlength{\tabcolsep}{1.7mm}{
\begin{tabular}{ccccccccc}
\hline
Datasets   & Building Name & Building Type & Gross Floor Area & Data Points  & Mean    & Max  & Min     & Std   \\ \hline
Data 1   & Robin & College Dormitory &62893 & 8760  & 31.24  & 67.50 & 0.23  & 10.84\\
Data 2 & Fox  & Office  & 23392 & 8760 & 31.44 & 85.30    & 0.42 & 13.66\\ \hline
\end{tabular}}
\end{spacing}
\end{table}

\subsection{Parameter design}
The proposed PIF-Net framework consists of a two-stage data processing module and a building load forecasting module. For data processing, we apply the Local Outlier Factor (LOF) algorithm for anomaly detection and use the SVM-SHAP mechanism for interpretable feature selection to reduce dimensional redundancy. The forecasting module employs patch processing, a shared GRU network, and a gating mechanism to predict building loads. The hyperparameters and internal parameters of the framework are detailed in \textbf{Table \ref{tab3}} and \textbf{Table \ref{tab4}}, where (*) indicates the optimal values used in this study.

\begin{table}[!h]
\centering
\small
\begin{spacing}{1.2}
\caption{\textbf{Hyper-parameters of the PIF-Net framework}}
\label{tab3}
\small
\setlength{\tabcolsep}{5mm}{
\begin{tabular}{cccc}
\hline
& \textbf{Model}      & \textbf{Parameters} & \textbf{Value}   \\ \hline
\multirow{4}{*}{\textbf{Data Processing Module}} & \multirow{2}{*}{\textbf{LOF}}    &  k-neighbors   &  [5, 10*, 20, 30]     \\
 &                         & Contamination       & 0.05              \\ 
& \multirow{2}{*}{\textbf{ SVM-SHAP}}  & Kernel    &   RBF    \\
    &                   & Feature selection metric          &   SHAP value    \\    \hline
\multirow{7}{*}{\textbf{Forecasting Module}}      
& \multirow{7}{*}{\textbf{PIF-Net}}  & Look-back window   &[12, 24*, 48, 96]         \\
 &               & Time step feature   & Based on SVM-SHAP \\
 &                    & Target length       & 1          \\
&                      & Learning rate       & [0.0001, 0.001*, 0.01]          \\
 &                 & Optimizer   & Adam        \\
      &         & Max epoch           & [100, 200, 300*, 400]             \\
     &              & Batch size          &  [32, 64*, 128, 256]                \\ \hline
\end{tabular}}
\end{spacing}
\end{table}

\begin{table}[!h]
\centering
\small
\begin{spacing}{1.2}
\caption{\textbf{Inter-parameters of the PIF-Net model}}
\label{tab4}
\small
\setlength{\tabcolsep}{9mm}{
\begin{tabular}{ccclc}
\hline
  &  \textbf{Layer} & \multicolumn{2}{c}{\textbf{Parameters}} & \textbf{Value} \\ \hline
\multirow{8}{*}{\textbf{PIF-Net model}}
& \multirow{3}{*}{\textbf{Patching}}
& \multicolumn{2}{c}{Patch length}     & 4            \\
 &      & \multicolumn{2}{c}{Patch number}     & 6            \\
 &      & \multicolumn{2}{c}{Stride} & 2   \\  
  & \multirow{3}{*}{\textbf{GRU}}
& \multicolumn{2}{c}{Hidden-layer}     & 2            \\
 &      & \multicolumn{2}{c}{Hidden dimension} & 64   \\  
  &      & \multicolumn{2}{c}{Dropout} & 0.2  \\ 
   & \multirow{1}{*}{\textbf{Gating Mechanism}}
& \multicolumn{2}{c}{Activation function} & Softmax   \\ \hline

\end{tabular}}
\end{spacing}
\end{table}

\subsection{Evaluation indexes}
To evaluate the point prediction performance of the PIF-Net framework, we use seven standard metrics: Mean Absolute Error (MAE), Mean Squared Error (MSE), Root Mean Squared Error (RMSE), Mean Absolute Percentage Error (MAPE), Coefficient of Determination ($R^2$), Index of Agreement (IA), and Theil's U1 statistic. These metrics quantify the deviation and correlation between the predicted and actual building loads, defined as follows:
\begin{equation}
\text{MAE} = \frac{1}{n} \sum_{i=1}^{n} |y_i - \hat{y}_i|
\label{eq:mae}
\end{equation}

\begin{equation}
\text{MSE} = \frac{1}{n} \sum_{i=1}^{n} (y_i - \hat{y}_i)^2
\label{eq:mse}
\end{equation}

\begin{equation}
\text{RMSE} = \sqrt{\frac{1}{n} \sum_{i=1}^{n} (y_i - \hat{y}_i)^2}
\label{eq:rmse}
\end{equation}

\begin{equation}
\text{MAPE} = \frac{1}{n} \sum_{i=1}^{n} \left| \frac{y_i - \hat{y}_i}{y_i} \right| \times 100\%
\label{eq:mape}
\end{equation}

\begin{equation}
R^2 = 1 - \frac{\sum_{i=1}^{n} (y_i - \hat{y}_i)^2}{\sum_{i=1}^{n} (y_i - \bar{y})^2}
\label{eq:r2}
\end{equation}

\begin{equation}
\text{IA} = 1 - \frac{\sum_{i=1}^{n} (y_i - \hat{y}_i)^2}{\sum_{i=1}^{n} \left( |\hat{y}_i - \bar{y}| + |y_i - \bar{y}| \right)^2}
\label{eq:ia}
\end{equation}

\begin{equation}
\text{U1} = \frac{\sqrt{\frac{1}{n}\sum_{i=1}^{n}(y_i - \hat{y}_i)^2}}{\sqrt{\frac{1}{n}\sum_{i=1}^{n}y_i^2} + \sqrt{\frac{1}{n}\sum_{i=1}^{n}\hat{y}_i^2}}
\label{eq:u1}
\end{equation}
where $y_i$ and $\hat{y}_i$ represent the actual building load value and the predicted building load value at the $i$-th data point, respectively. $\bar{y}$ denotes the mean value of the actual building loads, and $n$ is the total number of samples in the testing set. Lower values of MSE, RMSE, MAPE, and U1 indicate smaller prediction errors, while $R^2$ and IA values closer to 1 signify a better fit and stronger agreement between the predictions and actual observations.

\subsection{Operation environment}
The basic operating environment of the proposed PIF-Net framework is Intel Core i9 14900HX CPU @ 5.8 GHz, 16 GB RAM, and NVIDIA GeForce RTX 5060. Our model is constructed using PyTorch with the assistance of PyCharm IDE.

\section{Numerical Verification}

This section evaluates the PIF-Net framework for building load forecasting through three experiments. We compare its performance against classic deep learning and recent long-sequence forecasting models using seven metrics: MAE, MSE, RMSE, MAPE, $R^2$, IA, and U1. Ablation studies isolate the contribution of individual components, and a sensitivity analysis examines the overall robustness of the framework.

\subsection{Comparison analysis with different models}

\begin{figure}[!h]
    \centering
    \captionsetup{labelfont=bf}
    \includegraphics[width=0.98\linewidth]{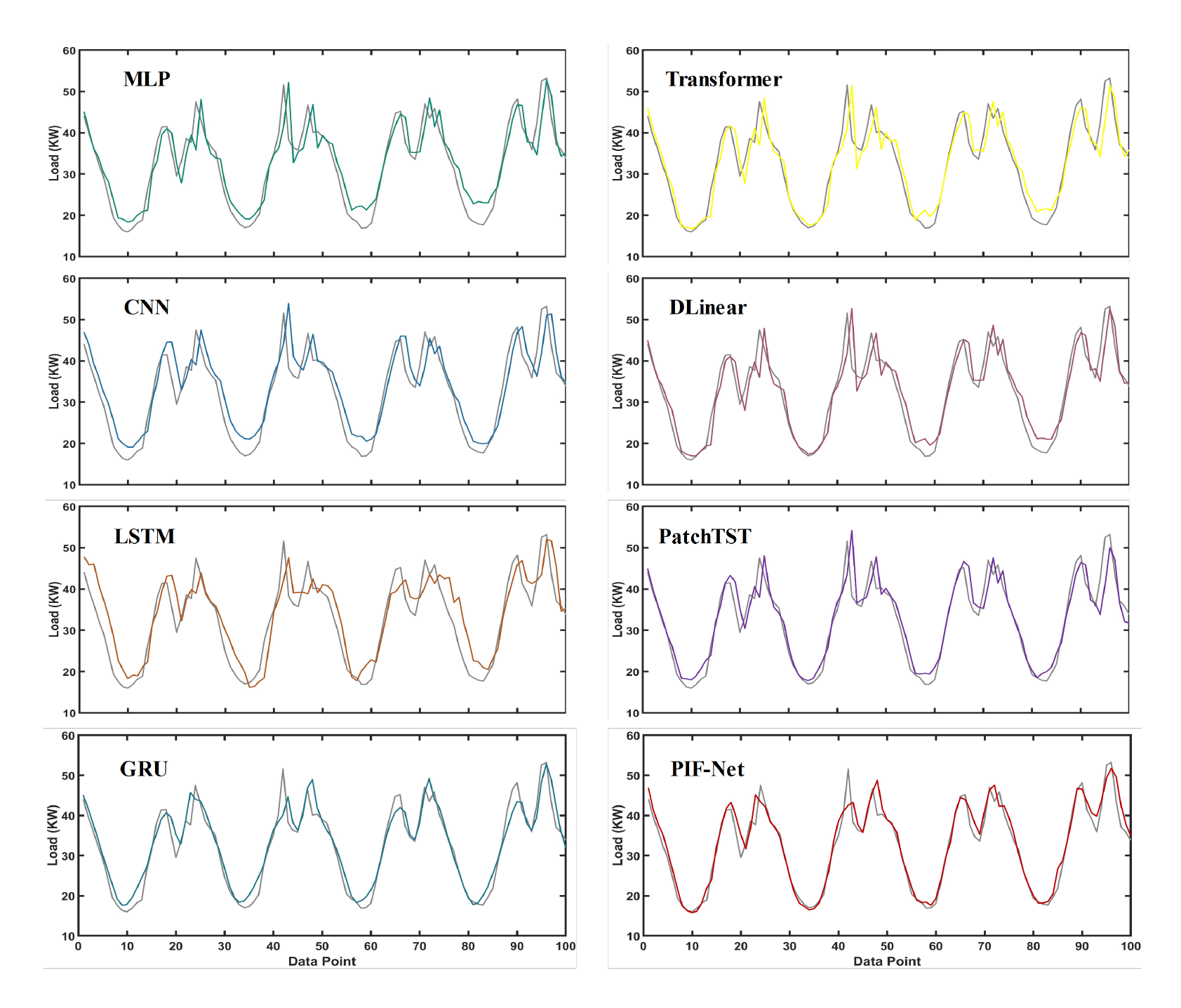}
    \vspace{-0.6\baselineskip}
    \caption{\textbf{Prediction results of different models on dataset 1}}
    \label{fig5}
\end{figure}

\begin{figure}[!h]
    \centering
    \captionsetup{labelfont=bf}
    \includegraphics[width=0.98\linewidth]{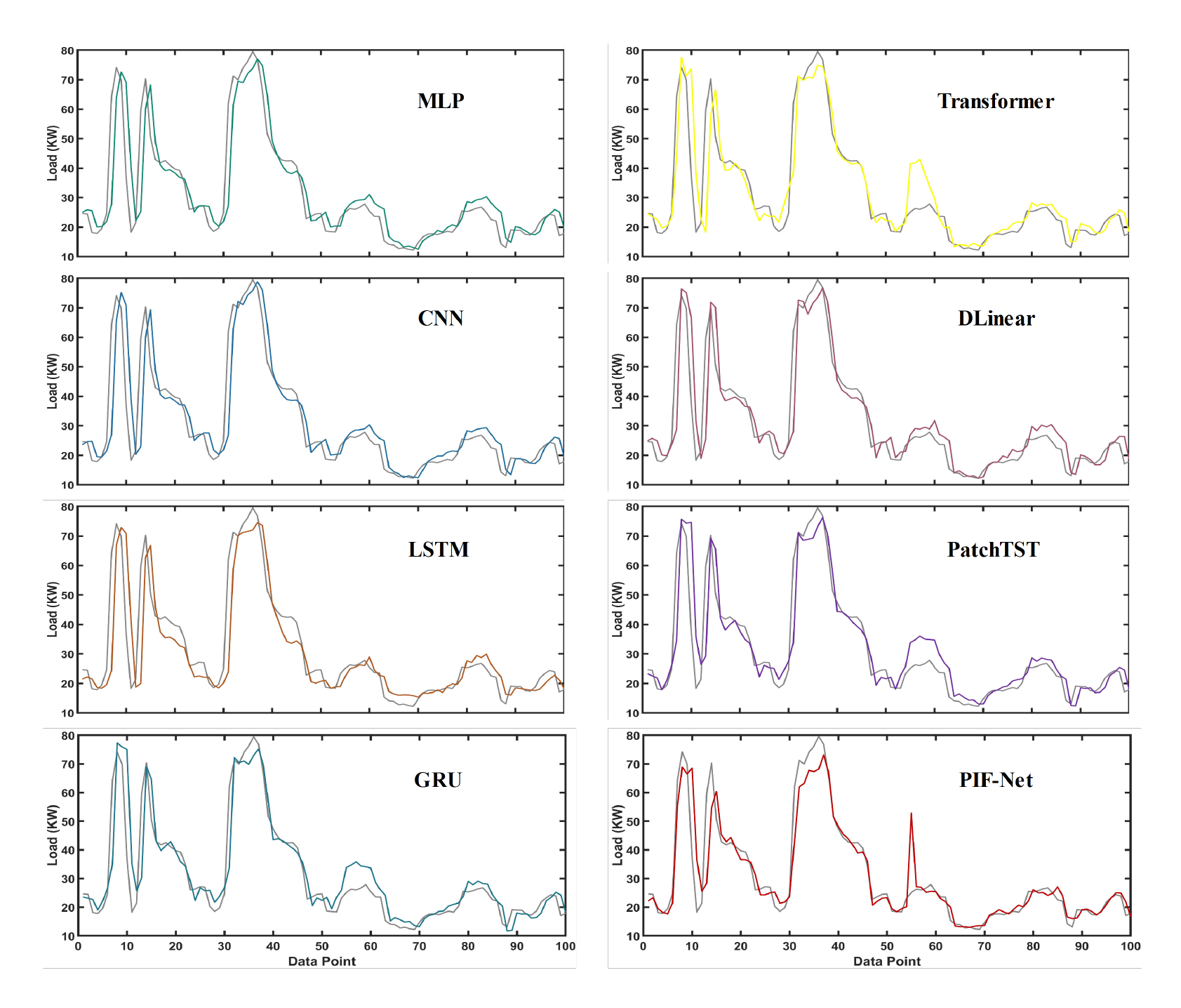}
    \vspace{-0.6\baselineskip}
    \caption{\textbf{Prediction results of different models on dataset 2}}
    \label{fig6}
\end{figure}

We compared the proposed PIF-Net against several classical deep learning models (MLP, CNN, LSTM, GRU) and recent long-sequence forecasting models (Transformer, DLinear, PatchTST). For a fair evaluation, all baseline models were trained using the same feature set, and their hyperparameters were fine-tuned for optimal performance. The forecasting results are shown in \textbf{Figure \ref{fig5}} and \textbf{Figure \ref{fig6}}, and the detailed quantitative results are presented in \textbf{Table \ref{tab5}}.

In Data 1, PIF-Net outperforms the baseline models across all evaluation metrics, comprehensively demonstrating its effectiveness and superiority. Compared to traditional models like CNN, PIF-Net decreased the MSE from 19.2335 to 6.2633 and the MAPE from 12.1497\% to 5.2775\%. Furthermore, it significantly increased the $R^2$ from 0.8136 to 0.9393, indicating the superior performance of PIF-Net in accurately capturing data variance. Against recent models like DLinear and PatchTST, PIF-Net maintains this performance gap. Compared to the patch-based model PatchTST, PIF-Net reduced MAE from 2.4486 to 1.7444 and RMSE from 3.5545 to 2.5027. Even against GRU, the best-performing baseline on Data 1, PIF-Net still achieved a notable reduction in MSE (from 8.8727 to 6.2633) and an improvement in $R^2$ (from 0.9140 to 0.9393).

In Data 2, which exhibits higher volatility and larger overall prediction errors, PIF-Net achieves an $R^2$ of 0.8742 and an IA of 0.9668, significantly higher than all comparison models, including MLP (0.7722), CNN (0.7666), LSTM (0.7449), GRU (0.8196), Transformer (0.7931), DLinear (0.8034), and PatchTST (0.8133). The experimental results indicate that PIF-Net can more accurately capture severe load variations and effectively fit the true values under complex conditions. Compared to LSTM, PIF-Net's MSE saw a massive reduction of 40.8225 (from 80.5597 to 39.7372), and the MAPE improved significantly from 15.3067\% to 12.6728\%. This strongly demonstrates its ability to mitigate large forecasting errors. Even when compared to the recent advanced PatchTST model, the advantages of PIF-Net remain pronounced. Specifically, PIF-Net decreased MSE by 19.2348 (from 58.9720 to 39.7372) and RMSE by 1.3756 (from 7.6793 to 6.3037), confirming its superiority over existing patch-based forecasting methods on this highly volatile dataset.

In conclusion, our model (PIF-Net) has demonstrated significant advantages in the experiments conducted on Data 1 and Data 2. Whether compared to classic deep learning models such as MLP, CNN, LSTM, and GRU, or advanced long-sequence prediction models like Transformer, DLinear, and PatchTST, PIF-Net consistently outperforms the comparison models across all evaluation metrics. The experimental results thoroughly showcase PIF-Net's powerful predictive capabilities in building load forecasting.

\textbf{Remark 1.} Experiment 1 validates the proposed end-to-end building load forecasting framework. By using multiple datasets with different building functional characteristics, the results confirm the model's generalization capability and performance consistency compared to baseline methods.

\newpage
\newgeometry{landscape, left=1.5cm, right=3cm,top=0.5cm,bottom=0.5cm}
\begin{landscape}
\pdfpagewidth=210mm 
\pdfpageheight=297mm 
\begin{table}[]
\begin{spacing}{1.4}
\caption{\textbf{Prediction results of two datasets in Experiment 1}}
\label{tab5}
\setlength{\tabcolsep}{4.5mm}
\begin{tabular}{lllllllll}
\hline
&  &\multicolumn{7}{l}{\textbf{Building load prediction}}\\ \hline
&  & \textbf{MAE}  & \textbf{MSE} & \textbf{RMSE} & \textbf{MAPE (\%)} & \textbf{R$^2$} & \textbf{IA} & \textbf{U1}  \\ \hline
\multirow{8}{*}{\textbf{Data 1}}   & MLP & 3.1354  & 15.7934 & 3.9741 & 10.7261  & 0.8469 & 0.9545 & 0.0585 \\ 
 & CNN & 3.5949  & 19.2335 & 4.3856 & 12.1497  & 0.8136 & 0.9471 & 0.0637 \\ 
 & LSTM & 4.0381  & 24.6663 & 4.9665 & 13.8937  & 0.7609 & 0.9328 & 0.0719 \\ 
 & GRU & 2.1766  & 8.8727 & 2.9787 & 7.2293  & 0.9140 & 0.9762 & 0.0437 \\ 
 & Transformer & 2.7473  & 13.8284 & 3.7187 & 8.4531  & 0.8660 & 0.9629 & 0.0550 \\ 
 & DLinear & 2.7289  & 13.7261 & 3.7049 & 8.5585  & 0.8670 & 0.9631 & 0.0547 \\ 
 & PatchTST & 2.4486  & 12.6346 & 3.5545 & 7.5944  & 0.8775 & 0.9663 & 0.0523 \\ 
& \textbf{Our Work (PIF-Net)}    & \textbf{1.7444}  & \textbf{6.2633} & \textbf{2.5027} & \textbf{5.2775 } & \textbf{0.9393} & \textbf{0.9849} & \textbf{0.0365}   \\ \hline
\multirow{8}{*}{\textbf{Data 2}}  & MLP & 4.6359  & 71.9303 & 8.4812 & 15.2996 & 0.7722 & 0.9362 & 0.1192 \\  
 & CNN & 4.5599  & 73.7198 & 8.5860 & 14.7126  & 0.7666 & 0.9370 & 0.1203 \\ 
 & LSTM & 4.8688  & 80.5597 & 8.9755 & 15.3067  & 0.7449 & 0.9296 & 0.1290 \\ 
 & GRU & 4.2040  & 56.9709 & 7.5479 & 14.7473  & 0.8196 & 0.9512 & 0.1052 \\ 
 & Transformer & 4.3259  & 65.3507 & 8.0840 & 14.8290  & 0.7931 & 0.9444 & 0.1128 \\ 
 & DLinear & 4.1998  & 62.0820 & 7.8792 & 13.8628  & 0.8034 & 0.9470 & 0.1103 \\ 
 & PatchTST & 4.3104  & 58.9720 & 7.6793 & 14.9006  & 0.8133 & 0.9491 & 0.1073 \\ 
& \textbf{Our Work (PIF-Net)}  & \textbf{3.4493}  & \textbf{39.7372} & \textbf{6.3037} & \textbf{12.6728 } & \textbf{0.8742} & \textbf{0.9668} & \textbf{0.0893} \\ \hline
\end{tabular}
\end{spacing}
\end{table}
\end{landscape}
\restoregeometry

\subsection{The ablation experiment of the PIF-Net framework}

\begin{figure}[!h]
    \centering
    \captionsetup{labelfont=bf}
    \includegraphics[width=1.0\linewidth]{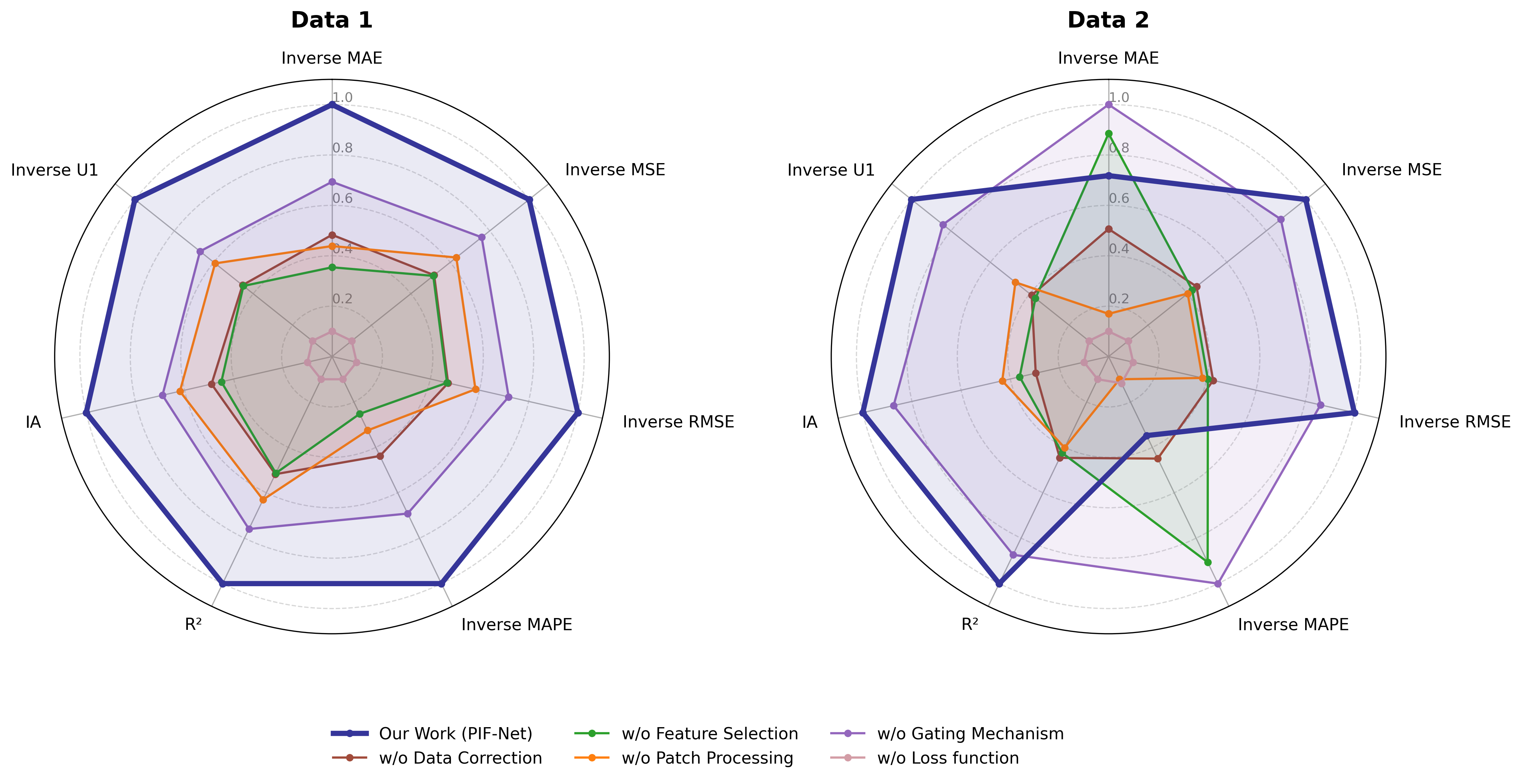}
    \vspace{-0.6\baselineskip}
    \caption{\textbf{Radar chart of performance comparison for ablation experiment}}
    \label{fig7}
\end{figure}

In this experiment, we conducted ablation studies to validate the necessity of each key component proposed in the PIF-Net framework across two different datasets. Specifically, we evaluated five variants of our model: removing the LOF-based data correction (w/o Data Correction), removing the SVM-SHAP feature selection (w/o Feature Selection), substituting the patching technology with standard sequential inputs (w/o Patch Processing), replacing the customized gating mechanism with standard average pooling (w/o Gating Mechanism), and replacing the proposed EWAL with the standard MSE loss function (w/o Loss function). \textbf{Table \ref{tab6}} demonstrates the experimental results in detail. Additionally, to facilitate a more intuitive comparison across multiple metrics, normalization was employed to scale the metric values into the range [0.1, 1], with the results visualized as a radar chart in \textbf{Figure \ref{fig7}}.

In Data 1, the ablation of any core component led to a distinct decline in the overall predictive performance of our PIF-Net framework. Notably, when the model was trained "w/o Loss function", we observed the most pronounced deterioration in performance. The MSE significantly increased by 8.4195 (from 6.2633 to 14.6828), and the $R^2$ decreased from 0.9393 to 0.8577. This demonstrates that the proposed EWAL is essential for handling the extreme volatility and enhancing the robustness of the model. Furthermore, the ablation of the two-stage data preprocessing also significantly weakened the model. When evaluated "w/o Feature Selection", the MSE worsened from 6.2633 to 10.8134, and the MAPE increased from 5.2775\% to 8.1123\%. Similarly, "w/o Data Correction" resulted in an MSE increase to 10.7714. In terms of the network architecture, removing the patch processing and the gating mechanism caused the $R^2$ to drop to 0.9058 and 0.9175, respectively. These results confirm that each module contributes to the final predictive accuracy of the framework on Data 1.

In Data 2, which exhibits higher complexity and baseline errors, the necessity of the proposed modules is further corroborated. When the model was trained "w/o Loss function", the performance declined drastically across all major metrics: the MSE surged by 19.6860 (from 39.7372 to 59.4232), the RMSE increased from 6.3037 to 7.7086, and the IA dropped from 0.9668 to 0.9503. This further confirms that EWAL's dynamic penalty mechanism is highly effective in stabilizing predictions under volatile conditions. Moreover, when the model was trained "w/o Patch Processing", the decline in temporal extraction capability was clearly reflected, with the MSE rising to 52.8333 and the $R^2$ dropping from 0.8742 to 0.8327. Likewise, the lack of feature redundancy filtering ("w/o Feature Selection") and local anomaly correction ("w/o Data Correction") caused the MSE to increase to 52.3333 and 51.8519, respectively. These results show that the two-stage preprocessing module, the patch-based gated network, and the adaptive loss function are all necessary for accurate building load forecasting. It is worth noting that although PIF-Net achieves the best MSE, RMSE, $R^2$, IA and U1, its MAE and MAPE are slightly higher than those of the model without the gating mechanism. This indicates that the proposed modules effectively capture peak loads and reduce the large prediction errors that dominate the MSE metric. However, this sensitivity to complex patterns introduces minor fluctuations during off-peak periods when actual loads are small, leading to an increase in MAPE. Because accurate peak load prediction is generally more critical for grid stability, this trade-off is acceptable.

\textbf{Remark 2.} Experiment 2 confirms the necessity of each component within the PIF-Net framework. The results show that combining LOF for data correction, SVM-SHAP for feature selection, a patch-based gating mechanism for spatiotemporal extraction, and EWAL for robust optimization is essential for accurate and reliable building load forecasting.

\newpage
\newgeometry{landscape, left=1.5cm, right=3cm,top=0.5cm,bottom=0.5cm}
\begin{landscape}
\pdfpagewidth=210mm 
\pdfpageheight=297mm 
\begin{table}[]
\begin{spacing}{1.4}
\caption{\textbf{Prediction results of two datasets in Experiment 2}}
\label{tab6}
\setlength{\tabcolsep}{4.5mm}
\begin{tabular}{lllllllll}
\hline
&  &\multicolumn{7}{l}{\textbf{Building load prediction}}\\ \hline
&  & \textbf{MAE}  & \textbf{MSE} & \textbf{RMSE} & \textbf{MAPE (\%)} & \textbf{R$^2$} & \textbf{IA} & \textbf{U1}  \\ \hline
\multirow{6}{*}{\textbf{Data 1}}   & w/o Data Correction & 2.4007  & 10.7714 & 3.2820 & 7.4091  & 0.8956 & 0.9734 & 0.0479 \\ 
 & w/o Feature Selection & 2.5644 & 10.8134 & 3.2884 & 8.1123  & 0.8952 & 0.9725 & 0.0480 \\  
 & w/o Patch Processing & 2.4573  & 9.7198 & 3.1177 & 7.8362  & 0.9058 & 0.9763 & 0.0450 \\ 
 & w/o Gating Mechanism & 2.1335  & 8.5156 & 2.9182 & 6.4471  & 0.9175 & 0.9779 & 0.0434 \\ 
 & w/o Loss function & 2.8872  & 14.6828 & 3.8318 & 8.6902  & 0.8577 & 0.9646 & 0.0553 \\ 
 & \textbf{Our Work (PIF-Net)} & \textbf{1.7444}  & \textbf{6.2633} & \textbf{2.5027} & \textbf{5.2775 } & \textbf{0.9393} & \textbf{0.9849} & \textbf{0.0365} \\ \hline
\multirow{6}{*}{\textbf{Data 2}}   & w/o Data Correction & 3.6292  & 51.8519 & 7.2008 & 12.2712  & 0.8358 & 0.9539 & 0.1019 \\ 
 & w/o Feature Selection & 3.3074  & 52.3333 & 7.2342 & 10.4654  & 0.8343 & 0.9551 & 0.1023 \\ 
 & w/o Patch Processing & 3.9154  & 52.8333 & 7.2687 & 13.6545  & 0.8327 & 0.9564 & 0.1002 \\ 
 & w/o Gating Mechanism & \textbf{3.2097}  & 42.4960 & 6.5189 & \textbf{10.0920}  & 0.8654 & 0.9645 & 0.0926 \\ 
 & w/o Loss function & 3.9745  & 59.4232 & 7.7086 & 13.5846  & 0.8118 & 0.9503 & 0.1079 \\ 
 & \textbf{Our Work (PIF-Net)} & 3.4493  & \textbf{39.7372} & \textbf{6.3037} & 12.6728  & \textbf{0.8742} & \textbf{0.9668} & \textbf{0.0893} \\ \hline
\end{tabular}
\end{spacing}
\end{table}
\end{landscape}
\restoregeometry

\subsection{Sensitive analysis}
In this experiment, we evaluate the robustness of the proposed PIF-Net framework through a hyperparameter sensitivity analysis. Unlike conventional deep learning models that heavily depend on strictly tuned hyperparameters to achieve optimal performance, PIF-Net is designed for high stability. By conducting a univariate sensitivity analysis, we aim to demonstrate that our model consistently maintains exceptional predictive accuracy without the need for exhaustive hyperparameter calibration.

Our sensitivity analysis focused on three core hyperparameters: Max Epoch (ME), Time Steps (TS), and Batch Size (BS). To isolate the impact of each factor, we systematically vary one hyperparameter while holding the others constant at their optimal values. The candidate sets are $\textbf{\textit{ME}}\in [100,\text{ }200,\text{ }{{300}^{*}},\text{ }400]$, $\textbf{\textit{TS}}\in [12,\text{ }{{24}^{*}},\text{ }{48},\text{ }96]$, and $\textbf{\textit{BS}}\in [32,\text{ }{{64}^{*},\text{ }128},\text{ }256]$, where the asterisk (*) designates the model's optimal configuration. Under this protocol, the notation $\textbf{\textit{ME}}\Phi (\textbf{\textit{TS}}, \textbf{\textit{BS}})$, for instance, denotes the scenario where only the Max Epoch is adjusted while Time Steps and Batch Size remain strictly fixed. To quantitatively evaluate the fluctuation in the model's predictive performance, we utilized the standard deviation (STD) as the sensitivity metric, which is mathematically defined as follows:

\begin{equation}
\text{\textit{STD}} = \sqrt{\frac{1}{n}\sum_{i=1}^{n}(y_i - \bar{y})^2},  
\end{equation}
where $y_i$  represents the performance metric values of the model under different hyperparameters, and $\bar{y}$ denotes the mean of $y_i$. We specifically conducted sensitivity analysis experiments on the building load forecasting task. The experimental results are detailed in \textbf{Table \ref{tab7}}.

\begin{table}[h]
\small
\begin{spacing}{1.2}
\caption{\textbf{Sensitive analysis result}}
\label{tab7}
\small
\setlength{\tabcolsep}{7mm}{
\begin{tabular}{lllll}
\hline
\multirow{2}{*}{}              & \multirow{2}{*}{\textbf{Index}} & \multicolumn{3}{l}{\textbf{PIF-Net}}       \\ \cline{3-5} &        & \textbf{$ME\Phi (TS,BS)$} & \textbf{$TS\Phi (ME,BS)$} & \textbf{$BS\Phi (TS,ME)$}  \\ \hline
\multirow{7}{*}{\textbf{Data 1}} & \textbf{MAE}   &0.2670  &0.2754    & 0.1599 \\
& \textbf{MSE}  &  1.3366 &1.7544  & 0.7664 \\
& \textbf{RMSE} &  0.2383 &0.2911  & 0.1431 \\
& \textbf{MAPE} &  0.0095 &0.0118 & 0.0058 \\
& \textbf{R$^2$} &  0.0130 &0.0172 & 0.0070 \\
& \textbf{IA}   &  0.0037 &0.0059  & 0.0025 \\
& \textbf{U1}   &  0.0038 &0.0046 &0.0028\\\hline
\multirow{7}{*}{\textbf{Data 2}} & \textbf{MAE}   &0.0968  &0.0768    & 0.1781 \\
& \textbf{MSE}  &  4.6462 &5.6186  & 6.4165 \\
& \textbf{RMSE} &  0.3251 &0.4207 & 0.4638 \\
& \textbf{MAPE} &  0.0035 &0.0018 & 0.0034\\
& \textbf{R$^2$} &  0.0147 &0.0178 & 0.0235 \\
& \textbf{IA}   &  0.0029 &0.0045  & 0.0052 \\
& \textbf{U1}   &  0.0038 &0.0052  &0.0071\\\hline
\end{tabular}}
\end{spacing}
\end{table}

The experimental results demonstrate that our proposed framework (PIF-Net) is relatively insensitive to hyperparameter variations. Across all tested configurations, the standard deviations for correlation metrics remain low, staying under 0.025 for $R^2$ and under 0.006 for IA on both datasets. Variations in relative error metrics are similarly constrained, with the STD of MAPE remaining below 0.012. These findings confirm the framework's robustness and predictive consistency across different parameter settings, supporting its practical application in building energy management.

\textbf{Remark 3.} Experiment 3 validates the stability of PIF-Net under various hyperparameter conditions, confirming that the model maintains robust performance without requiring exhaustive tuning.

\section{Conclusion}

In this study, we propose a highly robust and accurate end-to-end framework, PIF-Net, for building load forecasting. A two-stage data preprocessing module enhanced by interpretable feature selection first improves data quality. Specifically, the Local Outlier Factor (LOF) algorithm corrects anomalous spikes caused by sensor errors or extreme transient usage, and an SVM-SHAP mechanism evaluates feature importance to reduce dimensional redundancy. Within the forecasting module, patching segments long input sequences into localized temporal blocks to mitigate information loss. A shared Gated Recurrent Unit (GRU) network with residual connections extracts temporal features from each patch. A gating mechanism then assigns attention weights to these patches, aggregating them into a global representation for the final linear projection. To address optimization challenges from extreme load volatility, we propose an Error-Weighted Adaptive Loss (EWAL) function, which balances fitting precision and outlier tolerance to improve robustness. Evaluations on two distinct building datasets confirm that PIF-Net outperforms classic deep learning algorithms and recent long-sequence models.

Although the proposed PIF-Net framework exhibits exceptional forecasting capabilities, several areas remain for future work. Future research will focus on integrating broader multi-source data, such as real-time occupancy schedules, user behaviors, and spatial building topologies, to better model complex load dynamics. Subsequent investigations will also explore expanding this deterministic architecture into a probabilistic forecasting model to quantify predictive uncertainty, along with optimizing network compression techniques for efficient deployment on edge devices.

\section*{Acknowledgment}

This work was supported by the Research on Electricity Price Forecasting and Intelligent Bidding Strategy of Compressed Air Energy Storage in Qinghai Spot Market (Grant No. KFKT-26LAB-01).

\section*{Declaration of competing interest}

The authors declare that they have no known competing financial interests or personal relationships that could have appeared to influence the work reported in this paper.

\section*{Data availability}

We have utilized public datasets.

\bibliographystyle{elsarticle-num} 
\bibliography{cas-refs}

\end{document}